\documentclass[aps,10pt,twocolumn,prl,floats,floatfix]{revtex4-1}
\usepackage{verbatim}
\usepackage{listings}
\usepackage{graphicx}
\usepackage{color}
\usepackage{amsmath}
\usepackage{amssymb}
\usepackage[dvips]{epsfig}
\usepackage[T1]{fontenc}
\usepackage{shadow}
\usepackage{hyperref} 
\usepackage{varioref}
\usepackage{epsfig}
\usepackage{array}
\usepackage{rotating}
\usepackage{subfigure}
\usepackage{amsbsy}
\usepackage{todonotes}
\newcommand{\be}{\begin{equation}}
\newcommand{\ee}{\end{equation}}
\newcommand{\bs}{\begin{split}}  
\newcommand{\es}{\end{split}}     
\newcommand{\beqa}{\begin{eqnarray}}
\newcommand{\eeqa}{\end{eqnarray}}

\newcommand{\ir}{\mathbf E_{IR}^{(i)}} 
\newcommand{\hir}{\hat{\mathbf E}_{IR}^{(i)}}

\newcommand{\vE}{\mathbf E}
\newcommand{\cH}{\mathcal{H}}
\newcommand{\cP}{\mathcal{P}}
\newcommand{\vk}{\mathbf{k}}
\newcommand{\vp}{\mathbf{p}}

\newcommand{\la}{\left<}
\newcommand{\ra}{\right>}

\begin{document}
\title{Local Observables in a Landscape of Infrared Gauge Modes}

\author{Mikjel Thorsrud$^1$, David F. Mota$^1$, Federico R.\ Urban$^2$\\
$^1$ {\footnotesize{\it Institute of Theoretical Astrophysics, University of Oslo, P.O. Box 1029 Blindern, N-0315 Oslo, Norway}}\\
$^2$ {\footnotesize{\it Service de Physique Th\'eorique, Universit\'e Libre de Bruxelles}},\\
{\footnotesize{\it CP225, Boulevard du Triomphe, B-1050 Brussels, Belgium}}}
\noaffiliation


\begin{abstract} 
Cosmological local observables are at best statistically determined by the fundamental theory describing inflation.  When the scalar inflaton is coupled uniformly to a collection of subdominant massless gauge vectors, rotational invariance is obeyed locally.  However, the statistical isotropy of fluctuations is spontaneously broken by gauge modes whose wavelength exceed our causal horizon. This leads to a landscape picture where primordial correlators depend on the position of the observer. We compute the stochastic corrections to the curvature power spectrum, show the existence of a new local observable (the shape of the quadrupole), and constrain the theory using Planck limits.   
\end{abstract}

\maketitle

In the standard model of cosmology, inflation provides the initial conditions for the dynamics of the Cosmos: the quantum fluctuations generated during that time are believed to have seeded the primordial density perturbations leading to the structures we observe in the Universe today.  However, as observers today we are limited in our ability to test our hypotheses by causality, in that only what happened during the last 60 e-folds or so of inflation is directly accessible.  Had inflation lasted just a bit longer, the inflated size of the Universe would exponentially exceed our causal horizon, and the link between our observables and the parameters of the high energy theory of inflation would at best be statistical.  This is so because local observations do depend on the location (in space and time) of our observable bubble within the macro-bubble which inflation has produced.  The reason is the background of superhorizon infrared modes --- modes that at a given time are larger than the observer's horizon, but within the full inflated patch --- which eschew our sight and represent a space-time dependent, but locally (quasi) homogeneous, background upon which smaller wavelength modes develop.

Predictions for observables hence become a statistical problem --- having access to one and only one Universe the particular realisation we observe does not coincide with the average across the entire inflated patch --- with a mean and variance dictated by the parameters of the fundamental theory and those describing the observer (the size of the patch).  In simple models of adiabatic single scalar field inflation this ambiguity amounts to simply shifting the time coordinate, and the direct connection is preserved \cite{Unruh:1998}.  This however is not the case if isocurvature modes were present, or in generic multi-field scalar inflation \cite{Nelson:2012, Nurmi:2013, LoVerde:2013, LoVerde:2013b}.

Here we focus instead on canonical scenarios of scalar inflation, paired however with a collection of subdominant massless gauge vectors.  When the gauge fields are coupled directly to the inflaton, their dynamics are intertwined, and their fluctuations can become important creating detectable signatures in the Cosmic Microwave Background (CMB) which can be constrained experimentally \cite{Groeneboom:08, Hanson:09, Groeneboom:09,Pullen:2010, komatsu}.

We focus on the ghost-free example where the interaction Lagrangian   
\be
\mathcal{L}_{int}=-\frac{1}{4}\sum_{i=1}^n I^2(\phi) F^{(i)}_{\mu\nu} F^{(i)\mu\nu},
\label{lagrangian}
\ee
describes $n\ge3$ independent Abelian gauge vectors with field strength $F^{(i)}_{\mu\nu}=\partial_\mu A^{(i)}_\nu - \partial_\nu A^{(i)}_\mu$ coupled uniformly to the inflaton via a modulation function $I(\phi)$.  In particular, we will consider the case $\left<I\right> \propto a^{-2}$, with $a(\tau)$ the scale factor of the Friedmann-Lema\^itre-Robertson-Walker (FLRW) metric ($\tau$ is conformal time).  This choice is not arbitrary, as it corresponds to a background attractor solution for quite general classes of coupling functions $I(\phi)$ and inflaton potentials \cite{attractor1,attractor2,attractor3,attractor4,Yamamoto:multivector1}. Although vectors possess anisotropic stress, it is known that isotropy is attractive in this model ($n\ge3$ and uniform couplings), namely the vectors rearrange to produce an isotropic total energy momentum tensor \cite{Yamamoto:multivector1}. Consequently, imprints in the primordial density fluctuations created by the collection of background vectors (in the attractor configuration) respect rotational symmetry \cite{Yamamoto:multivector2}.  However, the coupling also produces a scale invariant spectrum of gauge modes which come on top of the background vector.  Since these infrared fluctuations originate from quantum mechanical fluctuations in a FLRW background, they are drawn from isotropic (and homogeneous) probability distributions.  Hence, a hypothetical super-observer with access to the entire inflated space would see a statistically isotropic universe. As causal observers, however, our observations are biased and we cannot expect local observables to obey the symmetries of the underlying theory. 

The physical significance of the bath of super horizon gauge modes is well appreciated in models with interactions of the type $I^2 F^2$ for their backreaction effects on the spacetime dynamics \cite{mukhanov} as well as for the corrections to primordial correlators \cite{peloso12}. The gauge kinetic coupling produces a nearly scale-invariant spectrum of frozen electric-type modes $\delta\vE(\vk)=-I(\phi) \partial_\tau \delta\mathbf A(\vk)/ a^2$ (the magnetic-type components decay rapidly).  A given quantum mode becomes classical (commuting) as its wavelength is stretched to superhorizon size.  The integral over modes which at a given time are superhorizon, i.e., $k<\mathcal H(\tau)$ with $\cH \equiv a H \equiv \partial_\tau a/a$ being the comoving Hubble parameter, adds up to a collection of classical infrared (IR) vectors $\ir(\tau)$. Such a vector appears homogeneous for a local observer limited by his causal horizon, pointing in a certain direction and with a constant magnitude over the accessible spatial patch.  Following \cite{peloso12} we model each IR vector as a Gaussian random field with variance
\begin{align}
&\left<0 \right| \ir \cdot \ir \left| 0\right> = 9H^4 N / 2\pi^2, \label{exp1}
\end{align} 
where $N$ is the number of e-folds from the start of inflation.  Here we have assumed absence of classical vector hairs at the start of inflation, namely that all gauge fluctuations originate from the Bunch-Davis vacuum.  We can picture the dynamics as a Gaussian random walk with a new piece (drawn from a distribution) with variance $9H^4 / 2\pi^2$ added to $\ir$ for each e-fold. Each IR vector thus performs a random walk in the space of all possible directions and with a mean norm scaling as $\big<|\ir|\big>\propto \sqrt N$.  The IR vectors appear homogeneous locally, but vary over distances much greater than the horizon; this leads to a landscape of associated signatures.  

There is a rich literature on the primordial vector imprints on the spectrum of density perturbations, see \cite{Maleknejad12,Soda:2012zm} 
and references therein. At the two-point level such corrections have been expressed in terms of an axially symmetric quadrupole parametrised by the amplitude $g$, its orientation in a given realisation being random, which can be constrained/probed experimentally \cite{Groeneboom:08, Hanson:09, Groeneboom:09,Pullen:2010, komatsu}. In our case the multitude of randomly aligned vectors creates instead a general quadrupole which can be described by 5 spherical harmonic coefficients $\{b_{2m}\}$. Out of these 5 degrees of freedom, 3 correspond to the random orientation on the sky, whereas the remaining 2 represent the intrinsic degrees of freedom we can make predictions for, namely the amplitude $g(k) \in \mathbb{R}$ and a new shape parameter $\chi \in [0,\pi/2]$ which is exactly scale invariant.  In \cite{longpaper} we show how to identify the Euler angles mapping arbitrary $\{b_{2m}\}$ to $\{ 0,0,\tilde b_{20},\tilde b_{21},0\} \Leftrightarrow \{g,\chi\}$ and prove its consistency: for each set $\{b_{2m}\}$ there is one and only one doublet $\{ g, \chi \}$.  The resulting anisotropic power spectrum depends not only on the scale $k=|\mathbf k|$, but also on the direction $\hat{\mathbf k}=(\sin\theta \cos\varphi, \sin\theta \sin\varphi, \cos\theta)$:
\be
\cP(\vk) = \cP_0(k)\!\left[1 + g(k)\! \left( \cos{\chi} A(\hat{\mathbf{k}}) + \sin{\chi} B(\hat{\mathbf{k}}) \right)\right],
\label{powerspectrum}
\ee
where the anisotropic functions are $A(\hat{\mathbf{k}})=\cos^2\theta-1/3$ and $B(\hat{\mathbf{k}})=\sin2\theta \cos\varphi / \sqrt3$.  The shape parameter measures the departure from axial symmetry; for $\chi=0$ there is one residual rotational symmetry and the power spectrum is invariant under $\varphi\rightarrow\varphi+\Delta\varphi$.

We are interested in the power spectrum of comoving curvature perturbations in spatially flat gauge  
\be
\la \zeta_\vk \zeta_\vp \ra \equiv 2(\pi^2/k^3) \delta^3(\vk + \vp) \cP(\vk). 
\ee
To proceed we split the power spectrum 
\be
\cP(\vk)=\cP_{\!\phi}(k) + \delta\cP(\vk),
\ee
where the two terms represent the contribution from the inflaton and the collection of vectors, respectively. The latter term can be further divided in an isotropic monopole part and an anisotropic quadrupole part.  We shall focus on the regime $\cP_{\!\phi} \gg \delta\cP$ which allows us to use the Planck best-fit value for the unperturbed part, $\cP_{\!\phi}=2.2\cdot10^{-9}$ \cite{planck}. In this regime the quadrupole part of $\delta\cP$, which is phenomenologically constrained by Planck to be maximum at the 2\% level relatively to $\cP_{\!\phi}$ \cite{komatsu}, grows proportionally to $\sqrt n$.  To ensure that the monopole part of $\delta\cP$ is suppressed w.r.t.\ $\cP_{\!\phi}$ as well, we derive the limit $nN_{ex} \ll 10^4$ (verified by Monte Carlo realisations), where $N_{ex}$ is the number of e-folds after the start of inflation when the mode $k=\mathcal{H}_0$ left the horizon (with $\mathcal{H}_0$ today's comoving Hubble parameter). In this regime it is also guaranteed that the energy density of the vectors is subdominant since $n\big< (\ir)^2 \big> \ll M_p^2H^2$ is equivalent to $nN_{ex}  \ll 10^9$. This allows us to neglect the backreaction on the spacetime dynamics and safely perform our calculations in a flat FLRW metric. We also emphasize that all results derived below rely on the existence of the isotropic background attractor solution mentioned above. Since the stability of this classical attractor under quantum corrections is questionable, we shall assume that the energy stored in the sum of quantum fluctuations is small compared to the energy in the zero-mode (background) vector,  $|\ir| \ll |\mathbf{E_0}^{\!\!\!\!(i)}|$. This is possible because $|\mathbf{E_0}^{\!\!\!\!(i)}|$ is related to the parameters of the underlying high-energy theory and therefore adjustable \cite{Yamamoto:multivector1} (whereas the stochastic build-up of $\ir$ only depends on the attractive $\left<I\right> \propto a^{-2}$ scaling and therefore not sensitive to the choice of these parameters). Contrarily to the single vector case ($n=1$), where the background component violates isotropy and there is a degeneracy between imprints created by $\ir$ and $\mathbf{E_0}^{\!\!\!\!(i)}$ \cite{longpaper}, in our case this allows us to study the unique signatures of $\ir$ in the safe regime $|\ir| \ll |\mathbf{E_0}^{\!\!\!\!(i)}|$.

In practice, the quadrupole correction to the power spectrum $\cP(\vk)$ for the curvature operator $\zeta_\vk$ is calculated via the in-in formalism. We ignore three-level corrections from background vectors since they only contribute to the monopole \cite{Yamamoto:multivector2}. $\ir$, on the other hand, create anisotropies in $\zeta_\vk$ via loop terms; the effective interaction Hamiltonian was first derived in \cite{peloso12} for $n=1$, and generalises in an obvious way for the multi-vector case.  Since different gauge vectors commute, we find that the correction reads
\be
\frac{\delta\cP(\vk)}{\cP_{\!\phi}} = \sum_{i=1}^n  \frac{24}{\epsilon} \frac{|\vE^{(i)}_{IR}(\tau_0)|^2}{3M_p^2H^2} N_k^2 \sin^2\theta_{\hat{\vk}(i)}, 
\label{manyv}
\ee
where $\epsilon = 1 - \partial_\tau{\cal H}/{\cal H}^2$ is the slow-roll parameter, $M_p$ is the reduced Planck mass, $\cos\theta_{\hat{\vk}(i)} = \hat{\mathbf k} \cdot \hir(\tau_0)$ and $N_k \in [0,N_{\cH_0}]$ is the remaining number of e-folds of inflation when the comoving mode $k$ crossed the horizon. Notice that it is the status of the multitude of IR vectors at the time $\tau_0\equiv -1/\mathcal{H}_0$ which dictates their imprint.  This is so because infrared gauge modes added to $\ir(\tau)$ after $\tau_0$ are inhomogeneous from our point of view: their contribution averages out to very good accuracy, see \cite{longpaper}.  Also note from Eq.\ (\ref{manyv}) that each vector gives an axisymmetric quadrupole correction with rotational symmetry along $\ir$. But the vectors originate from quantum fluctuations in a FLRW spacetime and their directions are random and uncorrelated. Thus the total quadrupole is not expected to be rotationally symmetric; in addition to the amplitude $g$ we need the new shape parameter $\chi$ to characterise the power spectrum.  Both parameters are stochastic and we computed their probability distributions by Monte Carlo realisations of the collection of Gaussian IR vectors.

\begin{figure}[htbp!]
\begin{center}
\includegraphics[width=.9\columnwidth]{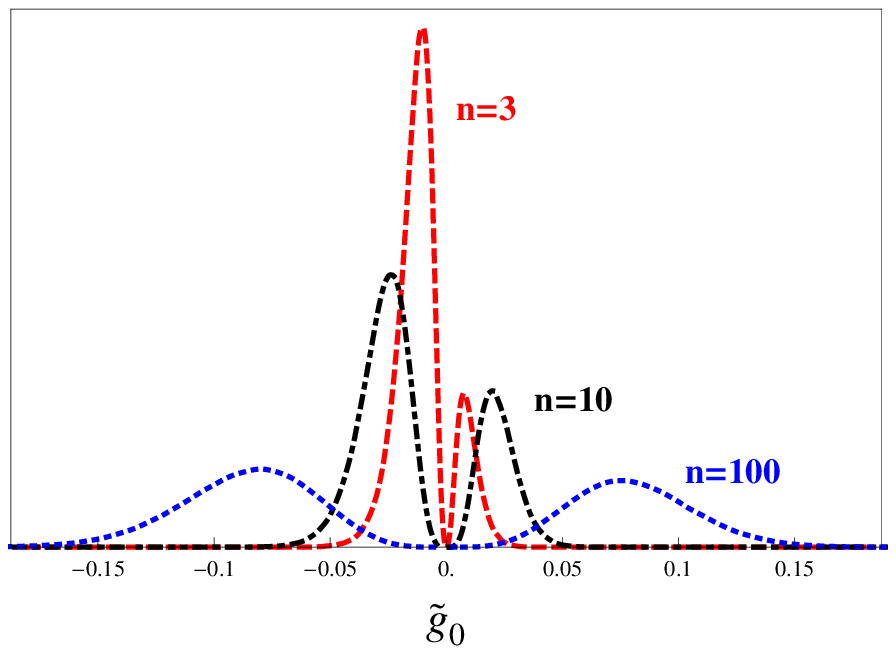}\par\vspace{10pt}
\includegraphics[width=.9\columnwidth]{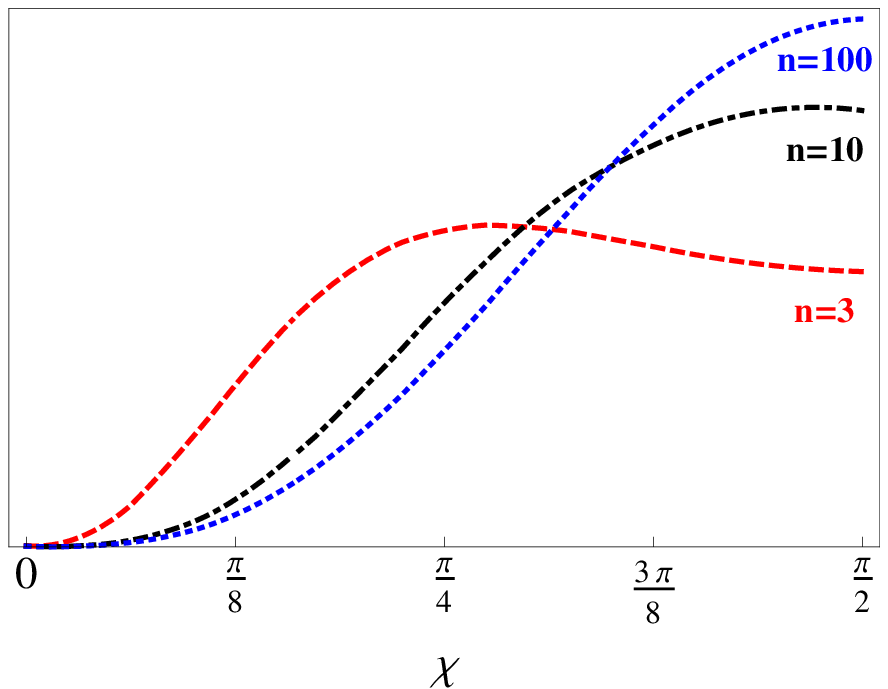}
\end{center}
\caption{Probability distributions for $\tilde g_0 \equiv g_0/(N_{ex}/3)$ (top panel) and $\chi$ (bottom panel) for $n=\{3,10,100\}$ gauge fields.}
\label{PDFs}
\end{figure}

In Fig.~\ref{PDFs} we show the probability distribution functions (PDF) for the amplitude $g_0 \equiv g(\mathcal{H}_0)$ and the shape $\chi$, for $10^6$ Monte Carlo realisations of $\hir(\tau_0)$, and for the three examples $n=3$, $n=10$, and $n=100$.  The amplitude has a logarithmic scale dependence, $g(k)\propto N_k^2$. For definiteness we set $N_{\cH_0}=60$  so that our reported values for $g_0$ roughly correspond to $k=\mathcal{H}_0$ in canonical models (this depends on the energy scale of inflation). With the normalisation $\tilde g_0 \equiv g_0/(N_{ex}/3)$ the PDF of $\tilde g_0$ only depends on the number of gauge fields; the corresponding physical parameter $g_0$ is obtained by picking a value for $N_{ex}$. The free parameter $N_{ex}$, which we constrain observationally below, can be thought of as the extra e-folds of inflation in addition to the 60 or so needed to solve the horizon problem. 

A local observer, clever enough to reconstruct the primordial power spectrum from cosmic data, sees a single random realisation of $g_0$ and $\chi$ drawn from these distributions. It is interesting to note that the statistics not only depend on the underlying theory ($n, N_\text{tot}$), but also on the size of the patch the observer has access to.  The latter is encoded in the parameter $N_{ex}$ via a degeneracy with $N_\text{tot}$.  Let us write $N_{ex} =  N_\text{tot} - N_{\cH_0}$ where $N_\text{tot}$ is the total duration of inflation. While $N_\text{tot}$ belongs to the theory, $N_{\cH_0}$ reflects the size of the patch accessible to the observer: the latter can be specified given a concrete theory, around 60 in canonical cases for observers today.  Thus the only free parameter in addition to $n$ is $N_{ex}$, or equivalently $N_\text{tot}$. Below we shall constrain the parameter space $\{n,N_{ex}\}$ using Planck limits \cite{komatsu}.

\begin{figure}[htbp!]
\begin{center}
\includegraphics[width=.48\textwidth]{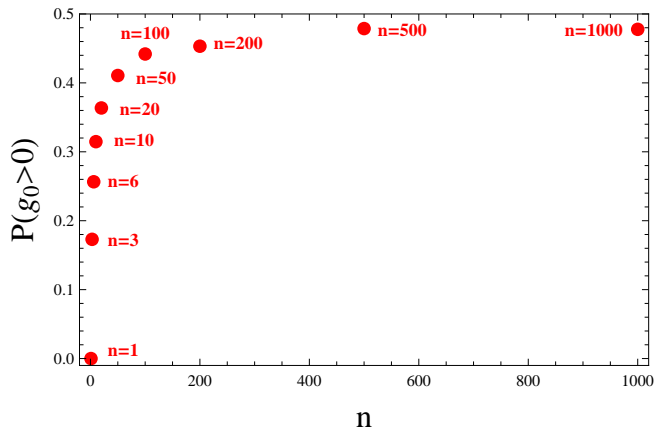}
\end{center}
\caption{Probability of obtaining a positive $g_0$ against the number of gauge fields $n$.}
\label{gprob}
\end{figure}

It is well-known that for a single field ($n=1$) in this setup one obtains a \emph{negative} value for the amplitude $g_0$.  We see how breaking the axial symmetry explicitly by introducing more vectors effectively results in a widening of the distribution, which is expected, but the distribution also become more symmetric in negative and positive values of $g_0$. Thus, the statistical nature of the $g_0$ parameter prevents us, single-bubble limited observers, from drawing a definite conclusion on its sign (and magnitude), even given the full dynamics at high energy known.  With a very large number of fields the probabilities of positive and negative amplitudes slowly approach: with $n=1000$ their values are about 0.48 and 0.52, respectively.  This trend is shown in Fig.~\ref{gprob}.

Single field $n=1$ models are characterised by an axial symmetric quadrupole which amounts to a vanishing shape parameter, $\chi=0$. The departure from axial symmetry is quantified by $\chi$, which is seen to grow with the number of fields.  The presence of the multitude of fields thus breaks explicitly the axial symmetry, the more so the more fields are thrown into the mix.

\begin{figure}[htbp!]
\begin{center}
\includegraphics[width=.48\textwidth]{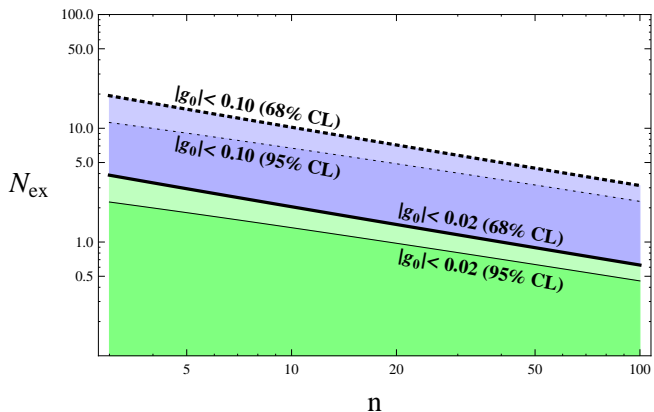}
\end{center}
\caption{Regions of parameter space for which $g_0$ is below 2\% and 10\% at 68\% CL and 95\% CL. The region below the bold continuous line is compatible with the Planck limit \cite{planck}.}
\label{Planck}
\end{figure}

We end the presentation of our results by showing the observationally allowed regions of the parameter space $\{ n, N_{ex} \}$ for which the generated quadrupole amplitude is at the 2\% and 10\% levels, relative to the monopole (Fig.~\ref{Planck}).  The lowest (green) area, below the $|g_0|<0.02 \; (95\%\text{CL})$ dashed line, is the region for which there is 95\% chance probability of generating $|g_0|<0.02$ (and a 5\% chance of having a higher value).  The following $|g_0|<0.02 \; (68\%\text{CL})$ solid line delimits the looser constraint (light green) for which there is a 68\% chance probability of generating a weaker than $|g_0| = 0.02$ quadrupole (and a 32\% chance of having a stronger one) --- this limit refers loosely to the recent analysis of \cite{komatsu} which placed the limit $-0.014 < g_0 < 0.018$ (68\% CL) from Planck data.  If we allow a 10\% quadrupole instead the parameter space widens encompassing the two middle (purple and light purple) regions.  In the uppermost (white) region the chance of finding a quadrupole below 0.10 decays to less than 68\%.

To summarise, in this Letter we studied observational implications of a collection of massless gauge vectors coupled uniformly to the inflaton.  Since the isotropic vector configuration is an attractor solution of the background equations, it was natural to focus on the case where both the background and perturbations respect rotational invariance.  In this setup we have shown how observations limited by a causal horizon are biased: local correlators are statistically expected to be anisotropic thereby violating the symmetry of the underlying model.  If inflation lasted only a few e-folds longer than the 60 or so needed to solve the horizon problem, the size of the entire inflated space exponentially exceeds that of our observable universe. This leads to a landscape picture where primordial correlators depend on the position of the observer. Predictions for observables thence become a statistical problem, with a mean and variance dictated by the parameters of the theory ($n$, $N_\text{tot}$), and those describing the observer ($N_{\cH_0}$).  This is a drastic theoretic leap from ignoring IR vector fluctuations where primordial correlators are deterministically dictated by the fundamental high energy theory. Such fluctuations were first taken into account quite recently, for $n=1$ \cite{peloso12}. In that case, however, the underlying model is fundamentally anisotropic and there is a degeneracy between imprints created by the background vector and those created by the IR gauge modes; there are no distinct signatures associated with the latter.

Conceptually there are striking parallels between our work and recent papers focusing on non-Gaussian landscapes in the multi-scalar or isocurvature contexts \cite{Nelson:2012, Nurmi:2013, LoVerde:2013, LoVerde:2013b}. In our vector setup, however, not only the statistics of already known observables are biased; remarkably, the very \emph{structure} of the two-point correlator is modified in a novel way leading to new types of locally detectable signatures.  Firstly, unlike single vector models ($n=1$) where the quadrupole correction can be described by a single amplitude $g_0$, there is in addition a new local observable described by the shape $\chi$, which is exactly scale invariant and measures the quadrupole's departure from axial symmetry. Secondly, and also unlike single vector models which predict negative values, it is possible with positive amplitudes as well; the probability for $g_0>0$ increases with the number of vectors and is close to $0.5$ for large $n$ (Fig.~\ref{gprob}). Both effects disappear if we disregard IR vector fluctuations \cite{Yamamoto:multivector2}.

To conclude, even modes whose wavelengths are much beyond what our causal observations can probe do have a distinct impact on the sky as we see it.  Despite our expectation the observable Universe being only a limited sample, we are still able to infer definite statistical predictions from within our Hubble Bubble. 

\paragraph{Acknowledgements} We would like to thank Nicola Bartolo, Sabino Matarrese and Angelo Ricciardone for very helpful discussions, comments and suggestions, Kei Yamamoto for reading through an early manuscript and giving valuable feedback, and Hans Kristian Eriksen and Dag Sverre Seljebotn for guidance on the experimental literature.

\bibliography{refs}

\end{document}